\documentclass[12pt,aps,showpacs,preprintnumbers,nofootinbib,tightenlines]{revtex4}

\usepackage{amsmath}
\usepackage{graphicx}
\usepackage[cp866]{inputenc}
\usepackage[english]{babel}

\newcommand{\tr}{\mathop{\rm tr}\nolimits}
\newcommand{\const}{\mathop{\rm const}\nolimits}
\newcommand{\be}{\begin{equation}}
\newcommand{\ee}{\end{equation}}
\newcommand{\ba}{\begin{aligned}}
\newcommand{\ea}{\end{aligned}}
\newcommand{\Pexp}{\mathop{\rm Pexp}\nolimits}
\newcommand{\inst}{\mathop{\rm inst}\nolimits}
\newcommand{\T}{\mathop{\rm T}\nolimits}
\newcommand{\NP}{\mathop{\rm NP}\nolimits}
\newcommand{\PP}{\mathop{\rm P}\nolimits}
\newcommand{\eff}{\mathop{\rm eff}\nolimits}
\newcommand{\dia}{\mathop{\rm dia}\nolimits}
\newcommand{\para}{\mathop{\rm para}\nolimits}
\newcommand{\cl}{\mathop{\rm cl}\nolimits}
\newcommand{\scc}{\mathop{\rm sc}\nolimits}
\newcommand{\sing}{\mathop{\rm sing}\nolimits}
\newcommand{\adj}{\mathop{\rm adj}\nolimits}
\newcommand{\cool}{\mathop{\rm cool}\nolimits}
\newcommand{\I}{\mathop{\rm I}\nolimits}
\newcommand{\fmfLc}[3]{\put(#1,#2){\raisebox{-0.5\height}{\makebox[0pt]{#3}}}}
\newcommand{\fmfLb}[3]{\put(#1,#2){\makebox[0pt]{#3}}}

\begin{document}

\title{Instantons in the nonperturbative QCD vacuum}
\author{N.O.~Agasian}
\email{agasian@heron.itep.ru}
\author{S.M.~Fedorov}
\email{fedorov@heron.itep.ru}
\affiliation{Institute of Theoretical and Experimental Physics,\\
117218, Moscow, B. Cheremushkinskaya 25, Russia}

\preprint{ITEP--PH--6/2001}

\begin{abstract}
The influence of nonperturbative fields on instantons in
quantum chromodynamics is studied. Nonperturbative vacuum is described in
terms of nonlocal gauge invariant vacuum averages of gluon field strength.
Effective action for instanton is derived in bilocal approximation and
it is demonstrated that stochastic background gluon fields are responsible
for infra-red (IR) stabilization of instantons.
Dependence of characteristic instanton size on gluon condensate and correlation
length in nonperturbative vacuum is found.
Comparison of obtained instanton size distribution with lattice data is made.
\end{abstract}


\date{17th September 2002}

\maketitle

\section{Introduction}
\label{sec_intr}

Instanton is the first explicit example of nonperturbative (NP) quantum
fluctuation of gluon field in QCD. It was introduced in 1975 by Polyakov
and coauthors~\cite{BPST}.
An important development originated with 't~Hooft's classic paper~\cite{tHooft_76},
in which he calculated
the semi-classical tunneling rate.
Instanton gas as a model of QCD vacuum was proposed in pioneer works by
Callan, Dashen and Gross~\cite{CDG_76,CDG_78}.
These topologically nontrivial field configurations
give a possible explanation of some problems of quantum chromodynamics.
Instantons allow to explain anomalous breaking of $U(1)_A$ symmetry
and the $\eta'$ mass~\cite{tHooft_76b,Witten_Venez_79}.
Spontaneous chiral symmetry breaking (SCSB) can be explained with
the help of instanton and anti-instanton field
configurations in QCD vacuum~\cite{Diak_Pet_86}.
An important role of instantons in scalar and pseudoscalar
channels was pointed out in paper~\cite{GI}. Taking into account
instantons is of crucial importance for many
phenomena of QCD (see~\cite{Scha_Shur_98} and references therein).

At the same time, there is a number of serious problems in instanton physics.
The first is the IR divergence of
integrals over instanton size $\rho$ at large $\rho$.
This makes it impossible to calculate instantons'
contribution to some physical quantities, such as vacuum gluon condensate.
Second, ''area law'' for Wilson loop can not be explained in instanton gas model,
hence quasiclassical instanton anti-instanton vacuum
lacks confinement which is responsible for hadron spectra.
It was recently demonstrated
that Casimir scaling which is observed in lattice calculations
for the interaction potential between heavy quarks $\bar{Q}Q$ in different
representations of  $SU(3)$ color group~\cite{Bali:1999hx}
can be violated in instanton gas~\cite{ShevSim}. This matter is also discussed
in~\cite{Fukushima:2000ix}.

There were many theoretical works aimed to solve the problem of
instanton instability in it's size $\rho$\footnote{Callan, Dashen and Gross suggested in their
pioneer works~\cite{CDG_78} to solve some of listed problems with the help of break-down of instantons
into meron-antimeron pairs.}. To some extent they all were based on an attempt to stabilize
instanton ensemble due to effects of interaction between pseudoparticles~\cite{Ilgenfritz:1980vj}.
Today the most popular is the model of ''instanton liquid'', which was phenomenologically
formulated by Shuryak~\cite{Shuryak_81,Shuryak:nr}. Using hypothesis that instantons dominate
in vacuum averages of local gludynamical operators, and taking into account dipole-dipole
interaction between instanton and antiinstanton, he determined main characteristics of the
medium: average distance between pseudoparticles is $\bar{R}\sim 1$~fm and their average
size is $\bar{\rho}\sim 1/3$~fm. Thus, a small quantity $\bar{\rho}/\bar{R}\simeq 1/3$ appears
and a picture of ''granular vacuum'' emerges, i.e. vacuum consists of well separated and
therefore not very much deformed instantons and anti-instantons. Nevertheless, action of pair
interaction $\exp\{S_{int}\}\gg 1$ appears to be important for the structure of instanton ensemble,
and therefore one can make a conclusion that instanton ensemble in QCD is not
a dilute gas, but an interacting liquid~\cite{Shuryak_81}.

Quantitatively similar results for parameters of instanton liquid were
obtained by Diakonov and Petrov~\cite{Diak_Pet_84}. They have suggested that vacuum
in gluodynamics consists of superposition of instantons and antiinstantons,
and found that stabilization is related to discovered by them classical repulsion between
pseudoparticles. However, further development~\cite{BY} revealed that instanton
ensemble can not be stabilized due to purely classical interaction. Thus, the mechanism
for the suppression of large-size instantons in the ensemble of topologically non-trivial
fields is still not understood.

A more natural assumption is that there exist other nonperturbative fields
apart from instantons in the vacuum, which allow to explain listed above problems. Interaction
of small-size instanton $\rho< 0.2$~fm with long-wave gluon fluctuations, described
by local vacuum condensate $\langle (g G^a_{\mu\nu})^2\rangle$, was studied in~\cite{SVZ}.
It was demonstrated that those fields lead to even faster growth of instanton density with
increasing of it's size $\rho$. On the other hand, in~\cite{Agas_Sim_95,Agasian_96} the study
of instantons in stochastic QCD vacuum, parametrized by nonlocal gauge invariant vacuum averages
of gluon field $\langle \tr G_{\mu\nu}(x)\Phi(x,y)G_{\sigma\lambda}(y)\Phi(y,x)\rangle$ (here $\Phi(x,y)$
is a parallel transporter) was started. It was shown that standard perturbation theory undergoes
a change in NP vacuum~\cite{Agas_Sim_95,Agasian_96} and that nonlocal interaction between large-size instanton
($\rho > 1 \mbox{fm}$) and nonperturbative background stops the infrared inflation~\cite{Agasian_96}.
Next, it was demonstrated in~\cite{AF_2001} that nonlocal interaction of instanton with NP
vacuum fields stabilizes it on scales of order of correlation length in vacuum
condensate.

In the present paper we develop gauge invariant method for calculating effective
action of an instanton in the NP vacuum. We show that instanton exists as a stable
topologically nontrivial field configuration with characteristic size $\rho_c$.
Value of $\rho_c$ functionally depends on properties of bilocal correlator
of NP field $\langle \tr G(x)\Phi(x,y)G(y)\Phi(y,x)\rangle$, i.e. it depends on two quantities:
$\langle G^2\rangle$ -- gluon condensate, and ''measure of vacuum inhomogeneity'' $T_g$ --
correlation length in NP vacuum. We don't discuss the problem of instantons distribution
in 4D-Euclidean space, i.e. the question about their density $N/V$. Consideration of this
problem in the framework of our approach requires studying interacting instanton-antiinstanton
ensemble against the background of nonperturbative fluctuations, described by gauge invariant
vacuum averages of gluon field.

This paper is organized as follows.
Section~\ref{sec_general} describes general formalism for effective action of an
instanton in nonperturbative background. In Section~\ref{sec_renorm} we study
one-loop renormalization of instanton action in the presence of NP fields. It is
shown that ''perturbative'' part of effective instanton action is ''freezed'' in
the NP background, and approaches constant value when instanton size is large enough
$\rho>m_*^{-1}\sim 1 \mbox{GeV}^{-1}$. Section~\ref{sec_IRstab} is devoted to the
study of ''direct'' interaction between instanton and NP fields. NP fields are
parametrized by gauge invariant nonlocal vacuum averages of gluon field (correlators).
Using cluster expansion we find effective action for an instanton in bilocal approximation
in a gauge invariant way.
Results of numerical calculations and discussion of results are given in Section~\ref{sec_numerical}.
Derivation of one-loop renormalization of effective instanton action in nonperturbative
vacuum in coordinate space, as well as some mathematical additions are brought out to Appendixes.

\section{General formalism}
\label{sec_general}
The influence of NP fluctuations on instanton can be separated into two parts.
First, perturbation theory undergoes a change in NP background, which results in the
change of standard one-loop renormalization of instanton action. Second, there appears
a direct nonlocal interaction of instanton with NP background field.

Standard Euclidean action of gluodynamics has the form
\begin{equation}
\label{eq_action}
S[A]=\frac{1}{2g_0^2} \int d^4 x \tr F_{\mu\nu}^2[A]=
\frac{1}{4} \int d^4 x F_{\mu\nu}^a[A]F_{\mu\nu}^a[A],
\end{equation}
where $ F_{\mu\nu}[A]=\partial_{\mu}A_{\nu} - \partial_{\nu}A_{\mu}-i[A_{\mu},A_{\nu}]$
is the strength of gluon field and we use the Hermitian matrix form for gauge fields
$A_\mu(x)= g_0A^a_\mu(x) {t^a}/{2}$ and $\tr t^at^b=\delta^{ab}/2$. We decompose $A_{\mu}$ as
\begin{equation}
\label{eq_decomp}
A_{\mu} = A_{\mu}^{\inst}+B_{\mu}+a_{\mu},
\end{equation}
where $A_{\mu}^{\inst}$ is an instanton-like field configuration with a unit topological charge
$Q_{\T}[A^{\inst}]=1$; $a_{\mu}$ is quantum field (expansion in $a_{\mu}$ reduces to
perturbation theory, which in gluodynamics leads to asymptotic freedom);
$B_{\mu}$ is nonperturbative background field (with zero topological charge), which can be
parametrized by gauge invariant nonlocal vacuum averages of gluon field
strength\footnote{In operator product expansion method and in QCD sum rules nonperturbative
field is characterized by a set of local gluon condensates $\langle G^2\rangle$,
$\langle G^3\rangle$, \ldots}.

In general case effective action for instanton in NP vacuum takes the form
\begin{equation}
\label{eq_genrl}
Z=e^{-S_{\eff}[A^{\inst}]} = \int
[Da_{\mu}]\left\langle e^{-S[A^{\inst}+B+a]}\right\rangle,
\end{equation}
where $\langle...\rangle$ implies averaging over background field $B_{\mu}$,
\begin{equation}
\label{eq_measure}
\left\langle\hat{O}(B)\right\rangle=\int d\mu(B) \hat{O}(B)
\end{equation}
and $d\mu(B)$ is the measure of integration over NP fields, explicit form of which is not
important for the following consideration.

Expanding $S[A]$ up to the second power in $a_{\mu}$ one obtains
$Z=\langle Z_1(B)Z_2(B)\rangle$, where
\begin{align}
&Z_1(B)=e^{-S[A^{\inst}]} \int
  [Da_{\mu}] \det (\nabla^2_{\mu}) \exp\left[\frac{1}{g^2_0}\tr\int d^4 x
  \left\{-(\nabla_{\mu} a_{\nu})^2+2iF_{\mu\nu}[a_{\mu},a_{\nu}]\right\}\right],
\label{eq_z1_z2_a}\\
&Z_2(B)=\exp\{-S[A^{\inst}+B]+S[A^{\inst}]\}.
\label{eq_z1_z2_b}
\end{align}
Here we use the notation $\bar{A}\equiv A^{\inst}+B$,
$F_{\mu\nu}\equiv F_{\mu\nu}[\bar{A}]$, and
$\nabla_{\mu}a_{\nu}=\partial_{\mu} a_{\nu}-i[\bar{A}_{\mu},a_{\nu}]$
is a covariant derivative.
Integration over $a$ and $B$ in~(\ref{eq_z1_z2_a}),(\ref{eq_z1_z2_b}) corresponds to
averaging over fields that are responsible for the physics at different scales.
Integration over $a_{\mu}$ takes into account perturbative gluons and
describes phenomena at small distances. Averaging over $B_{\mu}$ (formally
interaction with gluon condensate) accounts for
phenomena at scales of confinement radius. Therefore it is physically clear that averaging
factorizes and one obtains
\begin{equation}
\label{eq_factoriz}
Z \to \langle Z_1(B)\rangle \langle Z_2(B)\rangle.
\end{equation}
It should be noted that in the limit of infinite number of colors $N_c \to \infty$
factorization~(\ref{eq_factoriz}) is exact
$Z(N_c \to \infty) = \langle Z_1(B)\rangle \langle Z_2(B)\rangle$.
This allows us to write effective action of instanton in NP vacuum as a sum
of two terms, ''perturbative'' and ''nonperturbative'':
\begin{align}
\label{eq_seff_gnrl_a}
&S_{\eff}[A^{\inst}]=S^{\PP}_{\eff}[A^{\inst}]+S^{\NP}_{\eff}[A^{\inst}],\\
\label{eq_seff_gnrl_b}
&S^{\PP}_{\eff}[A^{\inst}]=-\ln\langle Z_1(B)\rangle,\\
\label{eq_seff_gnrl_c}
&S^{\NP}_{\eff}[A^{\inst}]=-\ln\langle Z_2(B)\rangle=
-\ln\left\langle \exp \{-S[A^{\inst}+B]+S[A^{\inst}]\}\right\rangle.
\end{align}

\section{One-loop renormalization of instanton action in NP vacuum}
\label{sec_renorm}

The general expression for one-instanton field configuration has the
well known form

\begin{equation}
\label{eq_inst}
A^{\inst}_{\mu} = 2 t^{b} R^{b \beta}
\overline{\eta}^{\beta}_{\mu\nu} \frac{(x-x_0)_{\nu}}{(x-x_0)^2}
f\left(\frac{(x-x_0)^{2}}{\rho^2}\right),
\end{equation}
where
\begin{equation}
\begin{aligned}
R^{b \beta} = 2\tr\bigl(\Omega t^{\beta}\Omega^{\dagger}t^b\bigr) \mbox{,\quad}
\Omega \in SU(N_c), \\
b = 1,2, \ldots N_c^2-1; \quad \beta = 1,2,3.
\end{aligned}
\end{equation}
$\overline{\eta}^{\alpha}_{\mu\nu}$ -- 't Hooft symbols, and matrix $R^{b \beta}$
ensures embedding of instanton into $SU(N_c)$ group, $R^{b\beta}$ satisfied relations

\begin{equation}
f^{abc}R^{b\beta}R^{c\gamma}=\varepsilon^{\beta\gamma\delta}R^{a\delta},\quad
R^{b\beta}R^{b\gamma}=\delta^{\beta\gamma},\quad
f^{abc}R^{a\alpha}R^{b\beta}R^{c\gamma} =\varepsilon^{\alpha\beta\gamma}.
\end{equation}

In singular gauge profile function
$f(z)$ satisfies boundary conditions $f(0)=1$, $f(\infty)=0$ and the
classical solution has the form

\begin{equation}
f(z^2)=\frac{1}{1+z^2}\,.
\label{eq_profile}
\end{equation}

The probability to find an instanton is determined by the classical
action functional $S_{\cl}[A]$ for the
solution~(\ref{eq_inst}),~(\ref{eq_profile}); that is

\begin{equation}
w\sim\exp\{-S_{\cl} [A^{\inst}_\mu]\}=\exp\{-8\pi^2/g^2_0\}\,.
\end{equation}

The preexponential factor was calculated in~\cite{tHooft_76}. The
result for one-instanton contribution to the QCD partition function is
$Z_{\I}= \int dn(\rho,x_0,R)$, where $dn$ is the differential
instanton density

\begin{align}
\label{eq_d0}
&dn(\rho,x_0,R)=[dR] d^4 x_0\frac{d\rho}{\rho^5} d_0(\rho),\\
&d_0(\rho)=\frac{4.6\exp\{-1.68 N_c\}}{\pi^2(N_c-1)!(N_c-2)!}
\left(\frac{8\pi^2}{g^2(\rho)}\right)^{2N_c}\exp \left
\{-\frac{8\pi^2}{g^2(\rho)}\right \}.
\end{align}

In the two-loop approximation in gluodynamics, the coupling constant
$g^2(\rho)$ is given by

\begin{equation}
\frac{8\pi^2}{g^2(\rho)}=
b\ln \left(\frac{1}{\rho\Lambda}\right) +\frac{b_1}{b}
\ln\ln\left(\frac{1}{\rho\Lambda}\right)+ O\left(\frac{1}{\ln(1/\rho\Lambda)}\right),
\qquad b=\frac{11}{3}N_c\,,\qquad b_1=\frac{17}{3} N_c,
\label{eq_coupling}
\end{equation}
$\Lambda\equiv\Lambda_{PV}$ corresponds to the Pauli-Villars regularization
scheme, $\Lambda\sim 200$\,MeV.

Callan, Dashen and Gross~\cite{CDG_78} were first to show that in a
constant gauge field instanton behaves as a colored four-dimensional
dipole. Shifman, Vainshtein and Zakharov~\cite{SVZ} generalized this
result to the case of interaction between a small-size instanton,
$\rho< 0.2$\,fm, and nonperturbative long-wave fluctuations that are
described by the local vacuum condensate $\langle G^2\rangle$. As a
result, $d_0 (\rho)$ is replaced by $d_{\eff}(\rho)$, where
\begin{equation}
d_{\eff} (\rho)\propto (\Lambda\rho)^{b}\left(1+\frac{4\pi^4\langle
G^2\rangle}{(N^2_c-1)g^4} \rho^4+\cdots\right).
\label{eq_d0_mod}
\end{equation}
Thus, we arrive at the well-known problem of the infrared inflation of
instantons in $\rho$.

Let us now consider how perturbative renormalization of effective instanton
action changes in NP vacuum~\cite{Agasian_96}. We will make use of background
field method, developed in~\cite{Abbot_81,Polyakov_87}, and stated in a
convenient way by Polyakov~\cite{Polyakov_87} when discussing asymptotic freedom
in nonabelian gauge theories.

In the absence of the term $\sim F_{\mu\nu}[a_{\mu},a_{\nu}]$,
there are four independent components of the field $a_{\mu}$ in
expression~(\ref{eq_z1_z2_a}) for $Z_1$.
Integration with respect to these independent
components yields $\left[\det^{-1/2}(\nabla^2_{\mu})\right]^4$.
Taking into account the ghost determinant in~(\ref{eq_z1_z2_a}),
we obtain $\left[\det^{-1/2}(\nabla^2_{\mu})\right]^2$. Thus,
there are all in all two physical polarizations for the field $a_{\mu}$
instead of four, as it must be. The last term in~(\ref{eq_z1_z2_a}),
$\sim F_{\mu\nu}[a_{\mu},a_{\nu}]$ describes the
interaction of the external field with the spin of the gluon
$a_{\mu}$.

Hence, there exist two effects. First, there is the
motion of charged particles in the external gluomagnetic field
$F_{\mu \nu}$. This motion leads to screening similar to
orbital Landau diamagnetism. Second, there is the direct
interaction of $F_{\mu \nu}$ with the spin of the field
$a_{\mu}$. In the second order in the external field, this
interaction yields an antiscreening term, which is analogous to
that describing the Pauli paramagnetic effect. With logarithmic
accuracy, these two phenomena can be separated. In the absence of
the term $\sim F_{\mu\nu}[a_{\mu},a_{\nu}]$ from~(\ref{eq_z1_z2_a}),
the expression for $S^{\PP}=-\ln Z_1$ then
completely corresponds to the effective action of massless scalar
electrodynamics with allowance for the isotopic factor. Hence, in
the second order in the background field, the ''diamagnetic''
contribution to $S^{\PP}$ can be written as

\be \label{eq_sdia_pt_1}
S^{\PP}_{\dia}=\int\frac{d^4q}{(2\pi)^4} \Pi^{ab}_{\mu\nu \dia}
(q)\bar A^a_\nu (q) \bar A^b_\nu(-q),
\ee
where $\Pi_{\dia}$ is given by the diagrams in Fig.~\ref{fig_1}.

\unitlength=1mm
\begin{figure}[!htb]
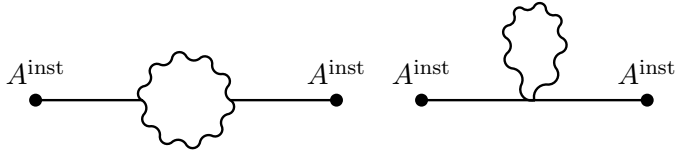

\begin{picture}(50,35)
\put(0,0){\includegraphics{diagr.2}}
\fmfLb{0}{14.60876}{{$A^{\inst}$}}%
\fmfLb{40}{14.60876}{{$A^{\inst}$}}%
\end{picture}
\begin{picture}(50,35)
\put(0,0){\includegraphics{diagr.3}}
\fmfLb{0}{14.60876}{{$A^{\inst}$}}%
\fmfLb{30}{14.60876}{{$A^{\inst}$}}%
\end{picture}
\caption {Diagrams for $\Pi_{\dia}$}
\label{fig_1}
\end{figure}
\unitlength=1pt

Let us first consider a simpler case of $\bar{A}=A^{\inst}$; that is, we set
the nonperturbative field $B_{\mu}=0$ and demonstrate how the standard perturbative
renormalization of the classical instanton action is reproduced. We then have

\be
\label{eq_dia_selfenrgy_1}
\Pi^{ab}_{\mu\nu \dia} (q) =2f^{acd}f^{bcd}\Pi^{(\scc)}_{\mu\nu}(q)=
2N_c\delta^{ab}\Pi^{(\scc)}_{\mu\nu}(q)
\ee

The factor of 2 in~(\ref{eq_dia_selfenrgy_1}) stems from two physical polarizations of
the field $a_{\mu}$. The expression for the polarization operator in massless scalar
electrodynamics has the form

\be
\label{eq_sc_selfenrgy}
\Pi^{(sc)}_{\mu\nu}(q)=-\frac14\int
\frac{d^4p}{(2\pi)^4}\frac{(2p+q)_\mu(2p+q)_\nu}{p^2(p+q)^2}
 + \frac12\delta_{\mu\nu}
\int\frac{d^4p}{(2\pi)^4}\frac{1}{p^2}= \Pi_1(q^2)(q^2\delta_{\mu\nu}-q_\mu q_\nu).
\ee
The well-known result for $\Pi_1$ in the leading logarithmic approximation is given by
\be
\label{eq_pi_log}
\Pi_1(q^2)=\frac{1}{192\pi^2}\ln \frac{\Lambda^2_0}{q^2},
\ee
where $\Lambda_0$ is the ultraviolet cutoff. Thus, we have
\be
\label{eq_dia_selfenrgy_2}
\Pi^{ab}_{\mu\nu \dia}(q) =2N_c \delta^{ab} (q^2\delta_{\mu\nu}-q_\mu q_\nu) \Pi_1(q^2)
\ee
Considering that the linear component $F_{\mu \nu}$ satisfies the relation
\be
\label{eq_f_relat}
F^a_{\mu\nu} (q)
F^{b}_{\mu\nu}(-q) =2(q^2\delta_{\mu\nu}-q_\mu q_\nu)\bar
A^a_\mu(q) \bar A^b_\nu(-q)
\ee
and
substituting~(\ref{eq_dia_selfenrgy_2}) and~(\ref{eq_f_relat}) into~(\ref{eq_sdia_pt_1}), we obtain
\be
\label{eq_sdia_pt_2}
\ba
&S^{\PP}_{\dia} =\frac14\int\frac{d^4 q}{(2\pi)^4}\Pi_{\dia}(q^2) F^a_{\mu\nu}(q)
  F^a_{\mu\nu}(-q), \\
&\Pi_{\dia} (q^2) =4N_c\Pi_1(q^2) =\frac{N_c}{48\pi^2} \ln
\frac{\Lambda^2_0}{q^2}
\ea
\ee

It should be noted that, in~(\ref{eq_sdia_pt_2}), the field strength involves
only the linear component $\sim\partial_{\mu}A_{\nu}$. However, from gauge invariance and
invariance under renormalization-group transformations, it follows that cubic
($\sim A^3$) and fourth-order ($\sim A^4$) terms appear in such a way that
they supplement $F^2_{\mu\nu}$ to the standard non-Abelian form.

Following Polyakov~\cite{Polyakov_87}, we represent the paramagnetic term in the form
\be
\label{eq_spara_pt}
\ba
&S^{\PP}_{\para} =-\frac{1}{2g^4_0}\int \frac{d^4q}{(2\pi)^4}  F^a_{\mu\nu} (q)
 F^b_{\lambda\rho}(-q)\times f^{acd} f^{bef} \langle a^c_\mu a^d_\nu (+q) a^c_\lambda
a^f_\rho(-q)\rangle \simeq\\
&\simeq -\frac{N_c}{4g^4_0} \int \frac{d^4q}{(2\pi)^4}
F^a_{\mu\nu}(q)
F^a_{\mu\nu}(-q) \int \frac{d^4q}{(2\pi)^4} \frac{2g^4_0}{p(p+q)^2}=\\
&=\frac14\int\frac{d^4q}{(2\pi)^4} \Pi_{\para} (q^2)
F^a_{\mu\nu}(q)  F^a_{\mu\nu}(-q),
\ea
\ee
where
\be
\Pi_{\para}(q^2)=-\frac{N_c}{4\pi^2} \ln
\frac{\Lambda^2_0}{q^2}
\ee
Bringing the above results together, we find that one-loop renormalized instanton action is given by
\be
\label{eq_seff_pt_1}
S^{\PP} =\frac{1}{4}\int
\frac{d^4 q}{(2\pi)^4} \left(\frac{1}{g^2_0}+\Pi (q^2)\right)
F^a_{\mu\nu}(q) F^a_{\mu\nu}(-q)
\ee
where
\be
\ba
&\Pi(q^2)=\Pi_{\dia}(q^2)+\Pi_{\para}(q^2)\\
&=\frac{N_c}{\pi^2}\left(\frac{1}{48}-\frac14\right)\ln
\frac{\Lambda^2_0}{q^2} =-\frac{11}{3}N_c
\frac{1}{16\pi^2}\ln\frac{\Lambda^2_0}{q^2}
\ea
\ee
As might have been expected, $S^{\PP}$ involves the effective
charge $g^2(q)$ defined at the external field momentum $q$
\be
\frac{1}{g^2(q)} =\frac{1}{g^2_0}-\frac{11}{3}
N_c\frac{1}{16\pi^2}\ln \frac{\Lambda^2_0}{q^2}
\ee
Using the standard renormalization-group arguments, we perform normalization at the momentum
\be \Lambda=\Lambda_0\exp
\{-\frac{8\pi^2}{b g^2(\Lambda_0)}\}
\ee
We then have
\be
\frac{1}{g^2(q)} =\frac{b}{16\pi^2} \ln
\frac{q^2}{\Lambda^2}, ~~ b=\frac13 N_c
\ee
We consider $S^{\PP}$ assuming that the background field is an
instanton. Because $F_{\mu\nu}^2$ is a gauge-invariant quantity, we use the regular gauge
\be
F^a_{\mu\nu}(x)=-4\eta^a_{\mu\nu}\frac{\rho^2}{(x^2+\rho^2)^2}
\ee
Considering that the Fourier transform in four-dimensional
Euclidean space is given by
\be
\tilde f(q)=\int d^4xe^{iqx}f(x)=\frac{4\pi^2}{q}\int^{\infty}_0
r^2drf(r)J_1(qr),~~ r=|x|
\ee
we obtain
\be
\label{eq_fstrngth_q}
\ba
&\tilde F^a_{\mu\nu}(q)=\int d^4xe^{iqx}   F^a_{\mu\nu}(x)\\
& =-16\pi^2\eta^a_{\mu\nu} \frac{\rho}{q} \int^\infty_0
\frac{r^2dr}{(r^2+\rho^2)^2}J_1(qr) =-8\pi^2\eta^a_{\mu\nu} \rho^2
K_0(\rho q)
\ea
\ee
where $K_0 (x)$ is a McDonald function. It can easily be shown
that the classical action for solution~(\ref{eq_fstrngth_q}) has the
standard instanton value
\be
\ba
&S_{\cl} [A^{\inst}]=\frac{1}{4g^2_0}\int\frac{d^4q}{(2\pi)^4} \tilde
F^a_{\mu\nu} (q) \tilde F^a_{\mu\nu} (-q)\\
&=\frac{8\pi^2}{g^2_0} 3\rho^4\int^\infty_0 q^3dqK^2_0(\rho q)
=\frac{8\pi^2}{g^2_0}\\
&\left (\int^\infty_0q^3dqK^2_0(\rho q)=\frac{1}{3\rho^4}\right)
\ea
\ee
Substituting~(\ref{eq_fstrngth_q}) into~(\ref{eq_seff_pt_1}), we arrive at
\be
\label{eq_seff_pt_2}
S^{\PP} [A^{\inst}]=8\pi^23\rho^4\int^\infty_0q^3dqK^2_0(\rho q)\frac{1}{g^2(q)}.
\ee
Let us introduce the dimensionless variable $x=\rho q$ and rewrite~(\ref{eq_seff_pt_2})
in the form
\be
\label{eq_seff_pt_3}
S^{\PP} [A^{\inst}]=8\pi^23\int^\infty_0
x^3dxK^2_0(x) \frac{1}{g^2(x/\rho)}
\ee
where
\be
\frac{1}{g^2(x/\rho)}=\frac{b}{16\pi^2}\ln
\frac{x^2}{\Lambda^2\rho^2}\\
=\frac{b}{16\pi^2}\ln \frac{1}{\Lambda^2\rho^2}
+\frac{b}{16\pi^2} \ln x^2
\ee
Thus, we finally obtain
\be
\label{eq_seff_pt_4}
S^{\PP} (\rho)
=\frac{8\pi^2}{g^2(\rho)}+\const = b\ln\frac{1}{\Lambda\rho}+\const.
\ee

We have reproduced the standard expression for $S^{\PP}(\rho)$. In the
instanton density $d(\rho)$, the constant from~(\ref{eq_seff_pt_4}) appears as a
preexponential factor, which is immaterial in our approach. We
will now study the modifications that arise in the above
equations as the result of taking nonperturbative vacuum
fluctuations into account.

It was shown in~\cite{Simonov_93_2} that, for the polarization operator in the
nonperturbative vacuum, the exponential decay of the form $\sim |x-y|^{-4}\exp\{-m_{*}|x-y|\}$ is
obtained in the infrared region instead of the standard power behavior
$\sim |x-y|^{-4}$ . Here $m_*\simeq 0.75 m_{0^{++}}  \sim 1\mbox{GeV}$, and $0^{++}$ is the lightest glueball
(see Appendix~\ref{appndx_altern}).
We will use this fact in analyzing instanton in the
nonperturbative vacuum. It can be seen from~(\ref{eq_seff_gnrl_b}) that
averaging is performed over the field B. This means that the
polarization operator in~(\ref{eq_seff_pt_1}) is represented by the diagrams in
Fig.~\ref{fig_2}.

Hence, in the expression~(\ref{eq_seff_pt_1}) for the effective action, we must
replace the perturbative expression for the polarization operator
by the polarization operator $\Pi_{\NP}(q^2)$ in the
nonperturbative background field. The latter was calculated in~\cite{Simonov_93_2}
and is given by

\be
\label{eq_propag_np}
\Pi_{\NP}(q^2)\simeq\frac{11}{3}N_c\frac{1}{16\pi^2}\ln
\frac{q^2+m^2_*}{\Lambda^2_0}
\ee

\unitlength=1mm
\begin{figure}[!htb]
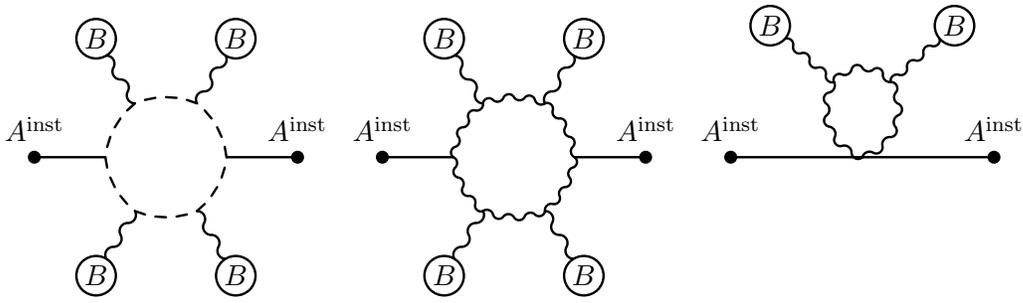

\begin{picture}(45,35)
\put(0,0){\includegraphics{diagr.4}}
\fmfLb{35}{19.60876}{{$A^{\inst}$}}%
\fmfLc{26.78406}{33.26558}{{$B$}}%
\fmfLc{8.21613}{33.26567}{{$B$}}%
\fmfLb{0}{19.609}{{$A^{\inst}$}}%
\fmfLc{8.21574}{1.73451}{{$B$}}%
\fmfLc{26.78368}{1.73424}{{$B$}}%
\end{picture}
\begin{picture}(45,35)
\put(0,0){\includegraphics{diagr.5}}
\fmfLb{35}{19.60876}{{$A^{\inst}$}}%
\fmfLc{26.78406}{33.26558}{{$B$}}%
\fmfLc{8.21613}{33.26567}{{$B$}}%
\fmfLb{0}{19.609}{{$A^{\inst}$}}%
\fmfLc{8.21574}{1.73451}{{$B$}}%
\fmfLc{26.78368}{1.73424}{{$B$}}%
\end{picture}
\begin{picture}(45,35)
\put(0,0){\includegraphics{diagr.6}}
\fmfLb{0}{19.60876}{{$A^{\inst}$}}%
\fmfLb{35}{19.60876}{{$A^{\inst}$}}%
\fmfLc{5.24979}{35}{{$B$}}%
\fmfLc{29.75021}{35}{{$B$}}%
\end{picture}
\caption{Diagrams for polarization operator $\Pi_{\NP}$ in the
nonperturbative background field}
\label{fig_2}
\end{figure}
\unitlength=1pt

Effective action~(\ref{eq_seff_pt_1}) is then written as
\be
\label{eq_seff_pt_5}
S^{\PP}_{\eff}[A^{\inst}]=\frac14\int \frac{d^4q}{(2\pi)^4}\tilde
F_{\mu\nu}(q) \tilde F^a_{\mu\nu}(-q) \frac{1}{g^2_{\NP}(q^2)},
\ee
where
\be
\frac{1}{g^2_{\NP}(q^2)}=\frac{b}{16\pi^2}\ln\frac{q^2+m^2_*}{\Lambda^2}.
\ee
Substituting the Fourier transform~(\ref{eq_fstrngth_q}) of the instanton field
into~(\ref{eq_seff_pt_5}) and introducing the variable $x=\rho q$, we obtain
\be
\label{eq_seff_pt_6}
\ba
&S^{\PP}_{\eff}[A^{\inst}]=24\pi^2\int^\infty_0x^3dxK^2_0(x)\frac{1}{g^2_{\NP}(x)}\\
&\frac{1}{g^2_{\NP} (x)}=\frac{b}{16\pi^2}\ln
\frac{x^2+m^2_{*}\rho^2}{\Lambda^2\rho^2}
\ea
\ee

Because $\ln$ is a slowly varying function and because the main
contribution to the integral in~(\ref{eq_seff_pt_6}) comes from the region
$x\sim 1$ ($K_0(x\gg1)\simeq \sqrt{\pi/2x}e^{-x}$), we can take $1/g^2_{\NP}(x)$ out of the
integral with logarithmic accuracy and set $x=1$. In this way, we
obtain~\cite{Agas_Sim_95,Agasian_96}
\be
\label{eq_seff_pt_final}
S^{\PP}_{\eff}(\rho)=\frac{8\pi^2}{g^2_{\NP}(1)}=\frac{b}{2} \ln
\frac{1/\rho^2+m_*^2}{\Lambda^2}
\ee

From~(\ref{eq_seff_pt_final}), it follows that the perturbative expression is
recovered for small-size instantons ($\rho\ll 1/m_{*}$) and that, for $\rho>1/m_{*}$, we
have $S^{\PP}_{\eff}\to \const$.

Derivation of formula~(\ref{eq_seff_pt_final}) in coordinate space is given in
Appendix~\ref{appndx_altern}.

\section{Interaction between instanton and nonperturbative
vacuum gluon fields}
\label{sec_IRstab}

We consider effect of NP fields on instanton, i.e. we evaluate $\langle
Z_2\rangle$.  In this work we make use of the method of vacuum
correlators, introduced in works of Dosch and Simonov~\cite{Dosch_87}.
NP vacuum of QCD is described in terms of gauge invariant vacuum averages of gluon fields (correlators)
$$
\Delta_{\mu_1 \nu_1 ... \mu_n \nu_n}=\langle\tr
G_{\mu_1\nu_1}(x_1) \Phi(x_1,x_2) G_{\mu_2\nu_2}(x_2) ...
G_{\mu_n\nu_n}(x_n) \Phi(x_n,x_1)\rangle,
$$
where $G_{\mu\nu}$ is gluon field strength, and
$\Phi(x,y)=\Pexp(i\int_y^x B_{\mu}dz_{\mu})$ is a
parallel transporter, which ensures gauge invariance. In many cases
bilocal approximation appears to be sufficient for qualitative description of
various physical phenomena in QCD.
Moreover, there are indications that corrections due to higher
correlators are small and amount to several percent~\cite{ShevSim,DShS}.

The most general form of bilocal correlator, which follows from
antisymmetry in Lorentz indices, is given by
\begin{align}
\label{eq_bilocal}
&\langle g^2 G_{\mu\nu}^a(x,x_0) G_{\rho\sigma}^b(y,x_0)\rangle
= \langle G^2 \rangle \frac{\delta^{ab}}{N_c^2-1} \times \notag\\
&\quad \times\left\{
\frac{D(z)}{12}(\delta_{\mu\rho}\delta_{\nu\sigma}-\delta_{\mu\sigma}\delta_{\nu\rho})
+\frac{\overline{D}(z)}{6}(n_{\mu}n_{\rho}\delta_{\nu\sigma}+n_{\nu}n_{\sigma}\delta_{\mu\rho}-
n_{\mu}n_{\sigma}\delta_{\nu\rho}-n_{\nu}n_{\rho}\delta_{\mu\sigma})\right\},
\end{align}
where
\begin{equation}
G_{\mu\nu}(x,x_0)=\Phi(x_0,x)G_{\mu\nu}(x)\Phi(x,x_0),
\end{equation}
$n_{\mu}={z_{\mu}}/{|z|}={(x-y)_{\mu}}/{|x-y|}$ is the unit vector,
$\langle G^2 \rangle \equiv \langle g^2 G_{\mu\nu}^a G_{\mu\nu}^a \rangle$
and, as it follows from normalization, $D(0)+\overline{D}(0)=1$.

Connection between functions $D(z)$, $\overline{D}(z)$ and standard functions
$D(z)$, $D_1(z)$, which are used in works~\cite{Dosch_87,DiGiacomo_2000,DiG,DShS},
is discussed in Appendix~\ref{appndx_bilocal}.

Functions $D(z)$ and $\overline{D}(z)$ have both perturbative and nonperturbative
contributions. We will consider only nonperturbative part, because perturbative one
has already been taken into account in $S^{\PP}_{\eff}$. Most data
about nonperturbative components of these functions come from numerical simulations on lattice.
Gluon condensate $\langle G^2 \rangle$ is also determined from lattice calculations,
but there exists a widely used estimate for this value based on charmonium spectrum analysis
and QCD sum rules $\langle G^2 \rangle \simeq 0.5\, \mbox{GeV}^4$~\cite{SVZ_79}.
According to the lattice data $D(z)$ and $\overline{D}(z)$ are exponentially decreasing functions
$D(z)=A_0 \exp(-z/T_g)$, $\overline{D}(z)=A_1 z \exp(-z/T_g)/T_g$, where $T_g$ is the
gluonic correlation length, which was measured on the lattice~\cite{DiG,BaliBrambillaVairo_98} and estimated
analytically~\cite{23} to be $T_g\sim 0.2$~fm.
Besides, according to lattice measurements $A_1 \ll A_0$ ($A_1 \sim A_0/10$).
Lattice data from paper of Di~Giacomo~\cite{DiGiacomo_2000} are presented in table~\ref{tab_digiacomo}.
$SU(3)$~full stands for chromodynamics with 4 quarks, while $SU(2)$ and $SU(3)$~quenched
stand for pure $SU(2)$ and $SU(3)$ gluodynamics, respectively. Bilocal correlator was also
measured on the lattice without using cooling method~\cite{BaliBrambillaVairo_98}, and correlation
length was found to be of around $0.1-0.2$ fm in quenched approximation.

Gluon condensate $\langle G^2 \rangle = 1.02 \pm 0.1 \mbox{GeV}^4$, obtained
by Narison~\cite{NN} basing on sum rules, is in good agreement with lattice
calculations~\cite{DiG}, $\langle G^2\rangle = 0.87$~GeV$^{4}$, for $SU(3)$ QCD.

\begin{table}[!htb]
\caption{Lattice data~\cite{DiGiacomo_2000} for correlation length}
\label{tab_digiacomo}
\begin{tabular}{lc}
\hline\hline
& $T_g$,~fm\\
\hline
$SU(2)$ quenched & 0.16 $\pm$ 0.02\\
$SU(3)$ quenched & 0.22 $\pm$ 0.02\\
$SU(3)$ full & 0.34\\
\hline\hline
\end{tabular}
\end{table}

To evaluate $S_{\eff}^{\NP}$ defined in~(\ref{eq_seff_gnrl_c}) we use the cluster expansion,
which is well known in statistical physics~\cite{IH}:
\begin{equation}
\label{eq_cluster}
\langle \exp(x) \rangle =
 \exp \left(\sum_n \frac{\langle\!\langle x^n \rangle\!\rangle}{n!} \right),
\end{equation}
where $\langle x \rangle = \langle\!\langle x \rangle\!\rangle$;~~~
$\langle x^2 \rangle = \langle\!\langle x^2 \rangle\!\rangle +
  \langle x \rangle^2$;~~~
$\langle x^3 \rangle = \langle\!\langle x^3 \rangle\!\rangle
  +3\langle x \rangle \langle\!\langle x^2 \rangle\!\rangle +
  \langle x \rangle^3$;~~~\ldots

\noindent One can easily find coefficients in cluster expansion, rewriting it in the following way
\be
\label{eq_cluster_coef}
 \begin{array}{rcl}
  \langle 1 \rangle &=& \langle\!\langle 1 \rangle\!\rangle  \\
  \langle 1\,2 \rangle &=& \langle\!\langle 1\,2 \rangle\!\rangle +
  \langle 1 \rangle\langle 2 \rangle \\
  \langle 1\,2\,3 \rangle &=& \langle\!\langle 1\,2\,3 \rangle\!\rangle +
  \langle 1 \rangle \langle\!\langle 2\,3 \rangle\!\rangle +
  \langle 2 \rangle \langle\!\langle 1\,3 \rangle\!\rangle +
  \langle 3 \rangle \langle\!\langle 1\,2 \rangle\!\rangle +
  \langle 1 \rangle \langle 2 \rangle \langle 3 \rangle \\
  .\,.\,.&&
\end{array}
\ee

Next, we modify expression~(\ref{eq_seff_gnrl_c}) for $S_{\eff}^{\NP}$
by adding $S[B]$ to the exponent:
\begin{equation}
S^{\NP}_{\eff}[A^{\inst}]=
-\ln\left\langle \exp \{-S[A^{\inst}+B]+S[A^{\inst}]+S[B]\}\right\rangle.
\end{equation}

In bilocal approximation this is equivalent to a change of
normalization of partition function,\footnote{Indeed, starting
from~(\ref{eq_seff_gnrl_c}) one finds that in bilocal
approximation~(\ref{eq_seff_4}) gets additional terms of the type
$\langle G^2 \rangle \int d^4 x$, which do not depend on $A^{\inst}$
and therefore result only in renormalization of $Z_2$.}  which is not
important for the following consideration.

Nonperturbative part of the effective instanton action
$S_{\eff}^{\NP}$ takes the form
\begin{align}
\label{eq_seff_3}
&S_{\eff}^{\NP}=\langle S[A^{\inst}+B]-S[B]-S[A^{\inst}]\rangle - \notag\\
&\frac{1}{2} \Bigl( \left\langle
(S[A^{\inst}+B]-S[B]-S[A^{\inst}])^2 \right\rangle -\left\langle
S[A^{\inst}+B]-S[B]-S[A^{\inst}] \right\rangle^2 \Bigr) + \ldots
\end{align}

We don't go further than
$\frac{1}{2}\langle\!\langle\bigl(S[A^{\inst}+B]-S[B]-S[A^{\inst}]\bigr)^2 \rangle\!\rangle$ term. Next terms of
cluster expansion decrease as $1/g^{2n}$. It is shown in~\cite{AF_2001} that $g^2(\rho_c) \sim 4$ on the
scale of instanton typical size, and therefore $1/g^2$ is a small parameter. Another small parameter
is $1/N_c$.

By using Fock-Schwinger gauge $x_{\mu}A_{\mu}=x_{\mu}B_{\mu}=0$
(instanton field $A_{\mu}^{\inst}$ satisfies it owing to the
properties of 't~Hooft symbols) and taking into account that
\begin{align}
\label{eq_Sab-sb}
S[A^{\inst}+B]-&S[B]-S[A^{\inst}] = \frac{1}{2g^2} \int d^4 x \tr \Bigl\{
- ([A^{\inst}_{\mu},B_{\nu}]-[A^{\inst}_{\nu},B_{\mu}])^2 \notag\\
&+2 F_{\mu\nu}[A^{\inst}]G_{\mu\nu}[B] - 4 i
(F_{\mu\nu}[A^{\inst}]+G_{\mu\nu}[B])[A^{\inst}_{\mu},B_{\nu}]
\Bigr\}
\end{align}
one gets
\begin{equation}
\label{eq_seff_4}
S_{\eff}^{\NP}=S_{\dia}+\frac{1}{2}S_{\dia}^2+S_{\para}+S_1+S_2,
\end{equation}
where in bilocal approximation
\begin{align}
&S_{\dia}=-\frac{1}{2g^2}\int d^4 x
\left\langle\tr\left[\left([A^{\inst}_{\mu},B_{\nu}]-[A^{\inst}_{\nu},B_{\mu}]
\right)^2\right]\right\rangle \label{eq_sdia},\\
&S_{\para}=-\frac{1}{2g^4}\int d^4 x d^4 y \left\langle
\tr\left(F^{\inst}_{\mu\nu}(x)G_{\mu\nu}(x)\right)
\tr\left(F^{\inst}_{\rho\sigma}(y)G_{\rho\sigma}(y)\right) \right\rangle \label{eq_spara},\\
&S_{1}=\frac{2}{g^4}\int d^4 x d^4 y \left\langle \tr\left(
F^{\inst}_{\mu\nu}[A^{\inst}_{\mu},B_{\nu}]\right)_x
\tr\left(
F^{\inst}_{\rho\sigma}[A^{\inst}_{\rho},B_{\sigma}]
\right)_y \right\rangle  \label{eq_s1},\\
&S_{2}=\frac{2i}{g^4}\int d^4 x d^4 y\left\langle
\tr\left(F^{\inst}_{\mu\nu}G_{\mu\nu}\right)_x
\tr\left(F^{\inst}_{\rho\sigma}[A^{\inst}_{\rho},B_{\sigma}]\right)_y
\right\rangle.
\label{eq_s2}
\end{align}

We use notations $S_{\dia}$ (diamagnetic) and $S_{\para}$
(paramagnetic) for contributions~(\ref{eq_sdia}) and~(\ref{eq_spara})
into interaction of instanton with background field. Physical
motivation for this is discussed in detail in~\cite{Agasian_96}.

We also take into account that
\be
\langle\tr (G_{\mu\nu} (A^{\inst}_{\mu}B_{\nu} - B_{\nu}A^{\inst}_{\mu})\rangle)
= i \frac{g^3}{2} f^{abc} \langle G_{\mu\nu}^a B_{\nu}^c\rangle
A_{\mu}^{\inst b} \propto f^{abc}\delta^{ac} = 0.
\ee

In Fock-Schwinger gauge
\be
\label{eq_fix_pnt}
B_{\mu}(x)=x_{\nu}\int\limits_0^1 \alpha d  \alpha
G_{\nu\mu}(\alpha x),
\ee
and thus we substitute vacuum averages by correlators in~(\ref{eq_sdia})-(\ref{eq_s2}). In general
form instanton field has the form
\be
\label{eq_inst_1}
A^{\inst}_{\mu}(x)=\Phi(x,x_0)A^{\sing}_{\mu}(x-x_0)\Phi(x_0,x).
\ee
Here $\Phi(x,y)=\Pexp\left(i\int\limits_y^x B_{\mu}dz_{\mu}\right)$ is the parallel transporter,
and $A^{\sing}_{\mu}(x-x_0)$ is instanton in fixed (singular) gauge~(\ref{eq_inst}),(\ref{eq_profile}).
We then obtain for field strength
\be
\label{eq_inst_f}
F^{\inst}_{\mu\nu}=\partial_{\mu}A^{\inst}_{\nu} -
\partial_{\nu}A^{\inst}_{\mu}-i[A^{\inst}_{\mu},A^{\inst}_{\nu}] =
\Phi(x,x_0)F^{\sing}_{\mu\nu}(x-x_0)\Phi(x_0,x)
+ \ldots\\
\ee
Dots denote terms which contain background field $B$, and therefore lead to higher
correlators~\cite{Agasian_96}, and we take into account bilocal correlator only.

Parallel transporters along with background field form gauge invariant
combinations, which are substituted by correlators after averaging. For instance, one
gets for $S_{\para}$
\be
\label{eq_s_para}
\ba
S_{\para}&=-\frac{1}{2g^4}\int d^4 x d^4 y \left\langle
\tr\left(F_{\mu\nu}(x)G_{\mu\nu}(x)\right)
\tr\left(F_{\rho\sigma}(y)G_{\rho\sigma}(y)\right) \right\rangle  = \\
&\ba
  =-\frac{1}{2g^4}\int d^4 x d^4 y \bigl\langle
    &\tr\left(\Phi(x,x_0) F^{\sing}_{\mu\nu}(x-x_0)
      \Phi(x_0,x)G_{\mu\nu}(x)\right) \times \\
    &\times \tr\left(\Phi(y,x_0)F^{\sing}_{\rho\sigma}(y-x_0)
      \Phi(x_0,y)G_{\rho\sigma}(y)\right)
    \bigr\rangle =
\ea\\
&=-\frac{1}{2g^4}\int d^4 x d^4 y \left\langle
  \tr\left(F^{\sing}_{\mu\nu}(x-x_0) G_{\mu\nu}(x,x_0)\right)
  \tr\left(F^{\sing}_{\rho\sigma}(y-x_0) G_{\rho\sigma}(y,y_0)\right)
  \right\rangle = \\
&=-\frac{1}{8g^2}\int d^4 x d^4 y
  \left(F^{\sing}_{\mu\nu}(x-x_0)\right)^{a}
  \left(F^{\sing}_{\rho\sigma}(y-x_0)\right)^{b}
  \left\langle g^2 G_{\mu\nu}^a (x,x_0)G_{\rho\sigma}^b (y,x_0) \right\rangle
\ea
\ee
Analogous calculations for other contributions to $S_{\eff}^{\NP}$~(\ref{eq_seff_4}),
lead to the following expressions (for the sake of simplicity we omit further
index 'sing' for instanton field in singular gauge, $A_{\mu}^{\sing}\equiv A_{\mu}$).

\begin{align}
&\ba
  S_{\dia}=&\frac{1}{2}
  \int d^4 x \int\limits_0^1 \alpha d \alpha \int\limits_0^1 \beta d\beta
  f^{abc}f^{dec}\bigl(
  A_{\mu}^a A_{\mu}^d x_{\lambda}x_{\rho}\langle g^2 G_{\lambda\nu}^b(\alpha x,x_0)
  G_{\rho\nu}^e(\beta x,x_0)\rangle \\
  -&A_{\mu}^a A_{\nu}^d x_{\lambda}x_{\rho}\langle g^2 G_{\lambda\nu}^b(\alpha x,x_0)
  G_{\rho\mu}^e(\beta x,x_0)\rangle
  \bigr)
  \ea
\\
&S_{\para}=-\frac{1}{8g^2}\int d^4 x d^4 y
F_{\mu\nu}^a(x)F_{\rho\sigma}^b(y)\langle g^2 G_{\mu\nu}^a(x,x_0)
G_{\rho\sigma}^b(y,x_0)\rangle\\
&\ba
    S_1=&-\frac{1}{2} \int d^4 x d^4 y
    F_{\mu\nu}^a(x) A_{\mu}^b(x) F_{\rho\sigma}^d(y) A_{\rho}^e(y) f^{abc}f^{def}\times\\
    &\times \int\limits_0^1 \alpha d \alpha \int\limits_0^1 \beta d\beta x_{\xi}y_{\eta}
  \langle g^2 G_{\xi\nu}^c(\alpha x,x_0) G_{\eta \sigma}^f(\beta y,x_0)\rangle
  \ea\\
&S_2=-\frac{1}{2g}\int d^4 x d^4 y \int\limits_0^1 \alpha d\alpha
F_{\mu\nu}^a(x) F_{\rho\sigma}^b(y) f^{bcd} A_{\rho}^c(y) y_{\xi}
\langle g^2 G_{\mu\nu}^a(x,x_0) G_{\xi\sigma}^d(\alpha y,x_0)\rangle
\end{align}

using representation~(\ref{eq_bilocal}) for gauge invariant condensate
$\langle g^2 G_{\mu\nu}^a(x,x_0) G_{\rho\sigma}^b(y,x_0)\rangle$ one
gets
\begin{align}
\label{eq_seff_component_2_a}
&S_{\dia} = \frac{\langle G^2 \rangle}{12} \frac{N_c}{N_c^2-1} \int
d^4 x \int\limits_0^1 \alpha d \alpha \int\limits_0^1 \beta d
\beta \, x^2 (A_{\mu}^a(x))^2 [D((\alpha-\beta)x)+2\overline{D}((\alpha-\beta)x)] \\
\label{eq_seff_component_2_b}
&\ba
  S_{\para}=&-\frac{\langle G^2 \rangle}{48 g^2} \frac{1}{N_c^2-1}
  \int d^4 x d^4 y \Bigl[F_{\mu\nu}^a(x) D(x-y) F_{\mu\nu}^a(y) +\\
  &+4\frac{(x-y)_{\mu}(x-y)_{\rho}}{(x-y)^2}F_{\mu\nu}^a(x)\overline{D}(x-y)F_{\rho\nu}^a(y)\Bigr]
\ea\\
\label{eq_seff_component_2_c}
&\ba
  S_{1}=&-\frac{\langle G^2 \rangle}{24}
  \frac{1}{N_c^2-1} \int d^4 x d^4 y \int\limits_0^1
  \alpha d \alpha \int\limits_0^1 \beta d \beta \,(xy
  \delta_{\nu\sigma}-x_{\sigma}y_{\nu})
  f^{abe}f^{cde}\\
  &\times F_{\mu\nu}^a(x)A_{\mu}^b(x)F_{\rho\sigma}^{c}(y)
  A_{\rho}^{d}(y)D(\alpha x - \beta y) + O(\overline{D})
\ea\\
\label{eq_seff_component_2_d}
&S_{2}=\frac{\langle G^2\rangle}{12 g}\frac{1}{N_c^2-1} \int d^4 x
d^4 y \,
f^{abc}F_{\rho\sigma}^a(y)A_{\rho}^b(y)F_{\sigma\nu}^c(x)y_{\nu}\int\limits_0^1
\alpha d \alpha D(x-\alpha y)+ O(\overline{D})
\end{align}

We have taken into account, that
\be
x_{\mu}\tilde B_{\mu}(x) \equiv x_{\mu}\left(\Phi(x_0,x)B_{\mu}(x)\Phi(x_0,x)\right)=
\Phi(x_0,x)\left( x_{\mu}B_{\mu}(x)\right) \Phi(x_0,x)=0,
\ee
and thus
\be
\tilde B_{\mu}(x) = x_{\nu}\int \alpha d \alpha \tilde G_{\mu\nu}(\alpha x) =
x_{\nu}\int\alpha d\alpha G_{\mu\nu}(\alpha x,x_0) + \ldots,
\ee
where in analogy with~(\ref{eq_inst_f})
\be
\tilde G_{\mu\nu}(x) =\partial_{\nu} \tilde B_{\mu} -
\partial_{\mu} \tilde B_{\nu} - i[\tilde B_{\mu},\tilde B_{\nu}]
= \Phi(x_0,x)G_{\mu\nu}(x)\Phi(x,x_0) + \ldots = G_{\mu\nu}(x,x_0) + \ldots
\ee

Thus we have obtained effective action for instanton in NP vacuum in
bilocal approximation\footnote{The tensor structure of instanton was
used to derive~(\ref{eq_seff_component_2_a})--(\ref{eq_seff_component_2_d}).}.
It can be seen from numerical calculation that typical instanton size
in QCD is $\rho_c\sim 0.25\div 0.3$\,fm.  For such $\rho$ instanton field in
it's center is strong $F_{\mu\nu}^2(x=x_0,
\rho_c)=192/\rho_c^4\gg\langle G^2\rangle$, and therefore classical
instanton solution is not much deformed in the region $|x|<\rho$, which
gives the main contribution to
integrals~(\ref{eq_seff_component_2_a})--(\ref{eq_seff_component_2_d}).

The problem of asymptotic behavior of instanton solution far from the
center $|x|\gg\rho$ was studied in detail in refs.~\cite{Diak_Pet_84,
MAK, Dor}. Our numerical analysis shows that the value of $\rho_c$ is
almost not affected by the asymptotic of classical instanton solution
provided that condensate $\langle G^2\rangle$ and correlation length
$T_g$ have reasonable values.

Next we have to define function $D(z)$. We use gaussian form
$D(z)=\exp({-\mu^2 z^2})$, $\mu\equiv{1}/{T_g}$. $D(z)$ is a
monotonously decreasing function with typical correlation length
$T_g$. Numerical calculations show that explicit form of
$D(z)$, as well as taking into account $\overline D(z)$ almost do not affect
the value of instanton size in NP vacuum $\rho_c$ (see Appendix~\ref{appndx_bar_d}).

We use the standard profile function
$f={\rho^2}/(\rho^2+x^2)$. Certainly, exact instanton profile (i.e.
profile that minimizes action) is different from this
one. Nevertheless, knowing the dependence of $S_{\eff}$ on $\rho$ we
will be able to make a conclusion about instanton behavior (inflation
or stabilization) and to determine it's typical size in NP vacuum.

Thus we obtain after integration over space directions
\begin{align}
\label{eq_component_4_a}
&S_{\dia}=\frac{2\pi^2}{g^2} \frac{\langle
G^2\rangle}{\mu^4}\frac{N_c}{N_c^2-1} \zeta^4
\int\limits_0^{\infty}dx
\frac{x^3}{(x^2+\zeta^2)^2} \varphi(x) \\
\label{eq_component_4_b}
&S_{\para}=\frac{-16\pi^4}{g^4}\frac{\langle
G^2\rangle}{\mu^4}\frac{1}{N_c^2-1} \zeta^4\int\limits_0^{\infty}
dx dy \frac{x^2 y^2
e^{-(x^2+y^2)}I_3(2xy)}{(x^2+\zeta^2)^2(y^2+\zeta^2)^2}\\
&\ba
  S_1=&\frac{-64\pi^4}{g^4} \frac{\langle
  G^2\rangle}{\mu^4}\frac{1}{N_c^2-1}
  \zeta^8\int\limits_0^{\infty} dx dy \frac{x^2
  y^2}{(x^2+\zeta^2)^3(y^2+\zeta^2)^3} \times \\
  &\times \int\limits_0^1 d\alpha \int\limits_0^1 d\beta
  e^{-(\alpha^2 x^2 + \beta^2 y^2)}(I_1(2\alpha\beta
  xy)+I_3(2\alpha\beta xy))
\ea\\
&S_2=\frac{-64\pi^4}{g^4} \frac{\langle
  G^2\rangle}{\mu^4}\frac{1}{N_c^2-1} \zeta^6\int\limits_0^{\infty} dx dy \frac{x^2
  y^2 e^{-x^2}}{(x^2+\zeta^2)^2(y^2+\zeta^2)^3}\int\limits_0^1 d\alpha
  e^{-\alpha^2 y^2} I_3(2\alpha xy)
\end{align}
where $I_n(x)=e^{-\frac{i\pi n}{2}}J_n(i x)$  is Bessel function of
imaginary argument (Infeld function),
$\zeta \equiv \mu \rho$ and
\begin{equation}
\label{eq_varphi}
\varphi(x)=\int\limits_0^1 \alpha d \alpha \int\limits_0^1 \beta
d \beta e^{-(\alpha-\beta)^2 x^2} =
e^{-x^2}\left(\frac{1}{3x^2}-\frac{1}{6x^4}\right) +
\frac{1}{6x^4} - \frac{1}{2x^2} + \frac{2}{3x} \Phi(x)
\end{equation}
where $\Phi(x)=\int\limits_0^x e^{-\xi^2}d\xi$ is error function.

We studied asymptotic behavior of instanton effective action and it's
dependence on dimensionless parameter $\zeta \equiv \mu \rho$. For a
small-size instanton $\rho\ll 1/\mu$ ($\zeta \rightarrow 0$) we found
for~$S_{\eff}^{\NP}$
\begin{align}
&S_{\dia}\to -\frac{\pi^2}{2 g^2}
\frac{N_c}{N_c^2-1}\langle G^2\rangle \rho^4 \ln(\mu\rho),\\
&S_{\para}+S_1 +S_2\to -\frac{2 \pi^4}{g^4}
\frac{\langle G^2\rangle}{N_c^2-1} \rho^4 +O\left(\langle G^2\rangle
\rho^4 (\mu \rho)\right).
\end{align}

In the opposite limit of large $\rho\gg 1/\mu$ ($\zeta \rightarrow
\infty$) one has
\begin{align}
&S_{\dia}\to \frac{\pi^{7/2}}{6 g^2}
\frac{N_c}{N_c^2-1}\langle G^2\rangle  \frac{\rho^3}{\mu},\\
&S_{\para}+S_1 +S_2\to \const.
\end{align}

Differential instanton density is proportional to ${dn}/ {d^4 z d\rho}
\propto \exp(-S_{\eff})$.  Thus, ``diamagnetic'' interaction of
instanton with NP fields leads to its infrared stabilization in
size~$\rho$.

\section{Discussion and Conclusions}
\label{sec_numerical}

In this work we have studied instanton behavior in NP vacuum, which
is parametrized by gauge invariant vacuum averages of gluon field
strength. We have derived effective instanton action in bilocal
approximation and have shown that ``diamagnetic'' term $S_{\dia}$
leads to IR stabilization of instanton.

\begin{figure}[!htb]
\begin{picture}(288,188)
\put(0,10){\includegraphics{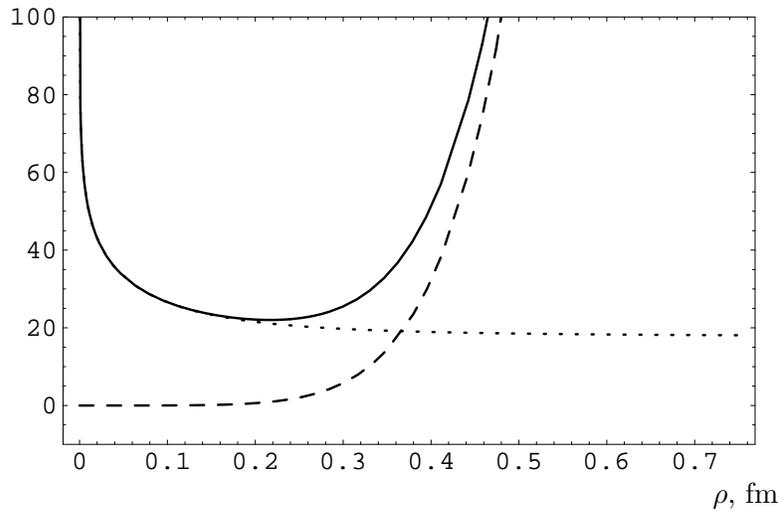}}
\put(270,0){$\rho$, fm}
\end{picture}
\caption{Effective action $S_{\eff}^{\PP}$ (dotted line),
$S^{\NP}_{\eff}$ (dashed line) and
$S_{\eff}=S_{\eff}^{\PP}+S^{\NP}_{\eff}$ (solid line) ($N_c=3$, $N_f=0$, $T_g=0.3$~fm, $\langle G^2 \rangle =
1.0$~GeV$^4$.)}
\label{fig_3}
\end{figure}

Numerical results for $S_{\eff}$ are plotted in Fig.~\ref{fig_3}. Three lines correspond to
$S^{\PP}_{\eff}(\rho)$, $S^{\NP}_{\eff}(\rho)$ and $S_{\eff}(\rho)=S^{\PP}_{\eff}(\rho)
+S^{\NP}_{\eff}(\rho)$  in the case of $SU(3)$ gluodynamics, at $\langle G^2\rangle = 1.0$~GeV$^4$
and $T_g=0.3$~fm.

In Table~\ref{tab_rho_c_nf0} values of $\rho_c$ for different $\langle G^2 \rangle$ and $T_g$
in case of $SU(3)$ gluodynamics are listed. Values of $\rho_c$ in case of QCD with $N_f=2$ are
presented in Table~\ref{tab_rho_c_nf2}.

\begin{figure}[!htb]
\begin{picture}(288,188)
\put(0,10){\includegraphics{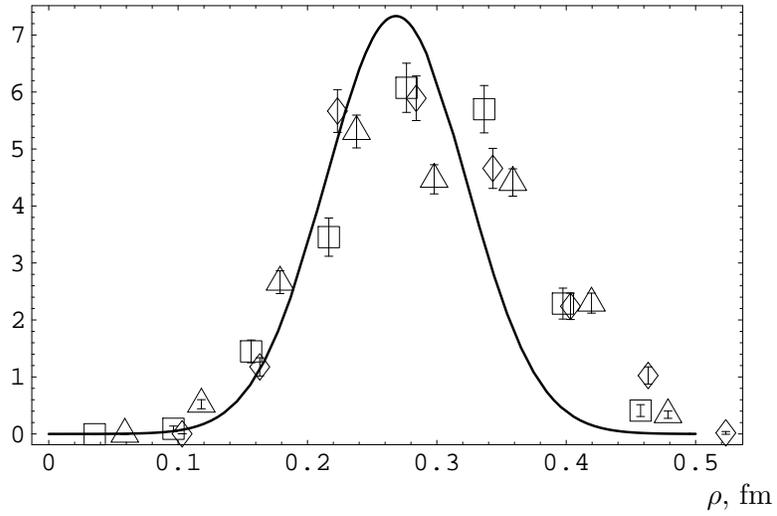}}
\put(270,0){$\rho$, fm}
\end{picture}
\caption{Instanton density $dn/d^4 z d\rho$ and lattice data~\cite{Hasen_Niet_98}
($N_c=3$, $N_f=0$, $T_g=0.2$~fm, $\langle G^2 \rangle = 0.5$~GeV$^4$.) }
\label{fig_4}
\end{figure}

Instanton differential density $dn/d^4 z d\rho \sim \exp(-S_{\eff})$ and
corresponding lattice data~\cite{Hasen_Niet_98} are presented in Fig.~\ref{fig_4}.
All graphs are normalized to the commonly accepted instanton density 1\,fm$^{-4}$.
Instanton size $\bar \rho \sim 1/3$~fm was first obtained in numerical calculations
on the lattice in~\cite{Polikarpov_87}. It should be mentioned that over the last years the lattice
results have not much converged (compare the
conferences~\cite{Latt_conf}).  Different groups roughly agree
on instantons size within a factor of two, e.g.\ $\bar{\rho}=0.3
,\ldots, 0.6$\,fm for $SU(3)$ gluodynamics.  There is no agreement at
all concerning the density $N/V$.  As a tendency, lattice studies give
higher density and larger instantons than phenomenologically assumed.

Thus, our results for $\rho_c$ are consistent with phenomenological value $\bar \rho$
and lattice data.
Furthermore, we can present physical arguments to explain
existing deviations with some lattice groups, which give larger than phenomenological
values of instanton size. Lattice calculations include cooling procedure, during
which some lattice configurations of gluon field are discarded.  This procedure can
result in a change in gluon condensate $\langle G^2\rangle$, and thus
instanton size distribution is calculated on the lattice at a value of gluon
condensate $\langle G^2\rangle_{\cool}$ which is different from physical
value $\langle G^2\rangle$. It should be noted that uncooled instanton size
distribution was studied in the work~\cite{Ringwald_99},
where authors introduced a new scaling variable (''cooling radius''), which
helps to extract information on the uncooled distribution.

\begin{figure}[!htb]
\begin{picture}(288,188)
\put(0,10){\includegraphics{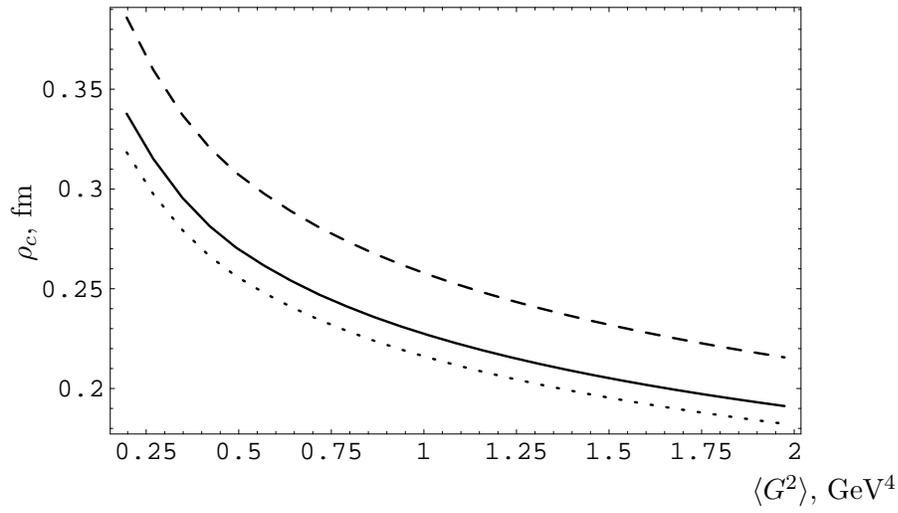}}
\put(270,0){$\langle G^2 \rangle$, GeV$^4$}
\put(-10,90){\rotatebox{90}{$\rho_c$, fm}}
\end{picture}
\caption{Instanton size as a function of gluon condensate ($N_c=3$, $N_f=0$)
         at $T_g=0.1$~fm (dashed line), $T_g=0.2$~fm (solid line),
         $T_g=0.3$~fm (dotted line)}
\label{fig_5}
\end{figure}
\begin{figure}[!htb]
\begin{picture}(288,188)
\put(0,10){\includegraphics{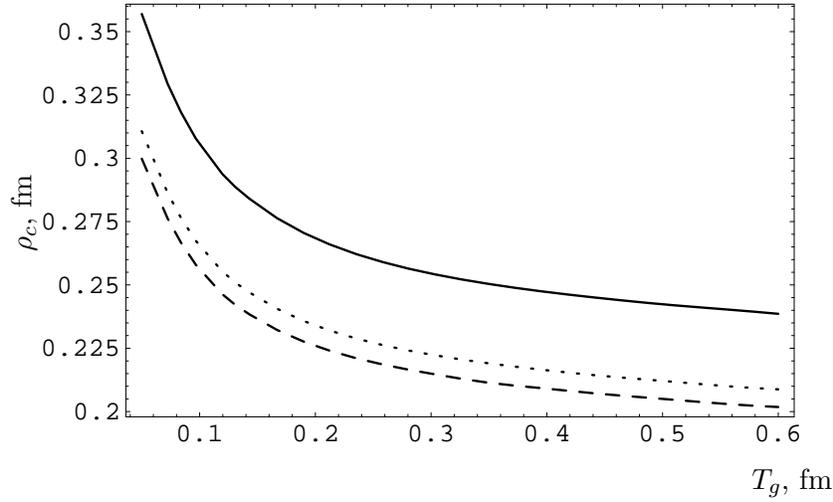}}
\put(-10,90){\rotatebox{90}{$\rho_c$, fm}}
\put(270,0){$T_g$, fm}
\end{picture}
\caption{Instanton size as a function of correlation radius ($N_c=3$, $N_f=0$) at
$\langle G^2 \rangle=0.5$~GeV$^4$ (solid line), $\langle G^2 \rangle=0.87$~GeV$^4$
(dotted line), $\langle G^2 \rangle=1.0$~GeV$^4$ (dashed line)}
\label{fig_6}
\end{figure}

We show dependence of $\rho_c$ on $\langle G^2\rangle$
for several values of $T_g$ in Fig.~\ref{fig_5}. One can see that
increase of $\langle G^2\rangle$ results in decrease of instanton
size, and that effect is a result of nonlocal ``diamagnetic''
interaction of instanton with NP fields. Fig.~\ref{fig_6} shows
$\rho_c$ as a function of $T_g$ for several values of $\langle
G^2\rangle$.  The smaller $T_g$, the larger instanton is. It is
physically clear that less correlated NP fields ($T_g \to 0$) have
smaller influence on instanton field configuration, which occupies 4D-Euclidean
volume with characteristic size $\rho$ ($\rho \gg T_g$).  On
the other hand, perturbative quantum fluctuations tend to inflate
instanton, and therefore $\rho_c$ increases with the decrease of
$T_g$.

We did not go beyond bilocal approximation in this work. As mentioned
above, this approximation is good enough not only for qualitative, but
also for quantitative description of some phenomena in nonperturbative QCD~\cite{DShS}.
In the problem under consideration there are two small parameters. These are
$1/g^2(\rho_c) \sim 0.15, \ldots, 0.25$ and $1/N_c$, and there powers
grow in each term of cluster expansion.
Moreover, one can make an estimate of the leading terms in cluster expansion.
It follows from our numerical calculations that in the region $\rho \gtrsim \rho_c$ dominating terms are
$S_{\dia}$, $\frac{1}{2} S_{\dia}^2$, $\frac{1}{3} S_{\dia}^3$ and so on. Summing them up, one gets
$S_{\dia} + \frac{1}{2} S_{\dia}^2 + \frac{1}{3} S_{\dia}^3 + \ldots = -\ln(1-S_{\dia})$.
Hence, taking into account next terms of cluster expansion not only conserves
IR stabilization, but even leads to some decrease of $\rho_c$. So, proposed
model describes physics of single instanton
stabilization in NP vacuum, not only qualitatively, but also
quantitatively with rather good accuracy.

\section*{Acknowledgements}

We are especially grateful to Yu.A.~Simonov for stimulating discussions and
useful remarks, and we thank E.-M.~Ilgenfritz, A.B.~Kaidalov, V.A.~Novikov and
A.V.~Yung for discussions of results.
The financial support of RFFI grant 00-02-17836 and INTAS grant N 110 is
gratefully acknowledged.

\newpage

\begin{table}[!htb]
\caption{$\rho_c$ (fm) for different values of $T_g$ and $\langle G^2 \rangle$, $N_c=3$, $N_f=0$}
\label{tab_rho_c_nf0}
\begin{tabular}{cc|cccc}
\hline
& $\langle G^2\rangle$, GeV$^4$ & 0.5 & 1.0 & 1.5 & 2.0\\
& $T_g$, fm \\
\hline
&0.1    &0.305      &0.256      &0.230      &0.213\\
&0.15   &0.282      &0.236      &0.213      &0.197\\
&0.2    &0.268      &0.226      &0.204      &0.189\\
&0.25   &0.256      &0.219      &0.198      &0.184\\
&0.3    &0.254      &0.215      &0.194      &0.181\\
&0.35   &0.250      &0.212      &0.191      &0.178\\
\hline
\end{tabular}
\end{table}

\begin{table}[!htb]
\caption{$\rho_c$ (fm) for different values of $T_g$ and $\langle G^2 \rangle$, $N_c=3$, $N_f=2$}
\label{tab_rho_c_nf2}
\begin{tabular}{cc|cccc}
\hline
& $\langle G^2\rangle$, GeV$^4$ & 0.5 & 1.0 & 1.5 & 2.0\\
& $T_g$, fm \\
\hline
&0.1    &0.308       &0.258       &0.232       &0.215\\
&0.15   &0.284       &0.238       &0.215       &0.199\\
&0.2    &0.270       &0.227       &0.205       &0.191\\
&0.25   &0.262       &0.221       &0.199       &0.185\\
&0.3    &0.256       &0.216       &0.195       &0.182\\
&0.35   &0.252       &0.213       &0.192       &0.179\\
\hline
\end{tabular}
\end{table}

\clearpage
\newpage

\appendix

\newpage
\section{Derivation of $S_{\eff}^{\PP}$ in coordinate space}
\label{appndx_altern}

Let us consider derivation of ~(\ref{eq_seff_pt_final}) in coordinate space~\cite{Agas_Sim_95}.

We start with expression~(\ref{eq_z1_z2_a}) for partition function, and normalizing it
to the case $A^{\inst}=0$ we obtain effective instanton action in nonperturbative
background
\begin{align}
\label{eq_seff_gnrl_alt_a}
&S_{\eff}[A^{\inst}]=S^{\PP}_{\eff}[A^{\inst}]+S^{\NP}_{\eff}[A^{\inst}]\\
\label{eq_seff_gnrl_alt_b}
&S^{\PP}_{\eff}=S_{\cl}-\ln \left\langle
  \frac{\det\nabla^2}{\det\nabla^2_0}\left(\frac{\det
  D^2}{\det D^2_0}\right)^{ 1/2}\right\rangle \\
\label{eq_seff_gnrl_alt_c}
&S^{\NP}_{\eff}=-\ln \left\langle \exp \{-S[A^{\inst}+B]+S[A^{\inst}]+S[B]\}\right\rangle
\end{align}
where
\be
\ba
&S_{\cl}=8\pi^2/g^2_0,\\
&\nabla^2=(\nabla_{\mu}[A^{\inst}+B])^2,\\
&(D^2)_{\mu\nu}=-\nabla^2\delta_{\mu\nu} +2iF_{\mu\nu}[A^{\inst}+B]
\ea
\ee
while subscript '0' implies that one should put $A^{\inst}=0$.

Keeping in~(\ref{eq_seff_gnrl_alt_b}) terms quadratic in instanton field and using~(\ref{eq_inst_1})
one has e.g. for the ghost contribution (gluon contribution can be obtained in the
same way, final result will contain, as always, two physical gluon polarizations)
\be
\label{eq_seff_pt_alt_1}
S^{\PP}_{\eff} = \int d^4x d^4y \tr A_{\mu}^{\sing} (x)
\partial_{\mu}\partial_{\nu} \tilde{\Pi} (x,y) A_{\nu}^{\sing} (y)
\ee
where explicitly gauge-invariant two-point function $\tilde{\Pi}$ is given by
\be
\label{eq_pi_tilde}
\tilde{\Pi}(x,y)=\langle\Phi_{\adj}(0,x)\nabla_0^{-2}(x,y)\Phi_{\adj}(y,0)
\nabla^{-2}_0(y,x)\rangle,
\ee
here $\nabla_0^{-2}(x,y)=\langle x| 1 / \nabla_0^2 |y\rangle$ is the ghost propagator.
One can distinguish in~(\ref{eq_pi_tilde}) a two-gluon correlator with an additional
condition that adjoint parallel transporter $\Phi_{\adj}$ is passing through the instanton
position $x_0=0$. The mass spectrum of $\tilde{\Pi}$ is bounded from below by the two-gluon
correlator without that condition $\Pi_0$, and the latter has as the lowest state $0^{++}$
glueball with the mass around 1.5 GeV. In the large-$N_c$ approximation $\Pi_0(Q)$
has only poles~\cite{Simonov_93_2}
\be
\label{eq_pi0_1}
\Pi_0(q^2)=\sum^{\infty}_{n=1} \frac{C_n}{q^2+M^2_n},~~M^2_n=4\pi
\sigma _{\adj}n+\const
\ee
as in $\bar{q}q$ case. The high-excited states in~(\ref{eq_pi0_1}) yield
\be
\label{eq_pi0_2}
\ba
&\Pi_0(q^2)\sim
\psi\left(\frac{q^2+m^2_0}{m_0^2}\right)+\const, \quad
\psi(x)=\frac{\Gamma'(x)}{\Gamma(x)}\\
&m^2_0=4\pi\sigma_{\adj}\simeq M^2(0^{++})
\ea
\ee
Replacing $\tilde{\Pi}$ by $\Pi_0$ in~(\ref{eq_seff_pt_alt_1})
and performing Fourier transform one has
\be
\label{eq_seff_pt_alt_2}
S^{\PP}_{\eff}= \frac{1}{4}\int \frac{d^4q}{(2\pi)^4}
\left(\frac{1}{g^2_0}+\Pi_0(q^2)\right)\tilde{F}^a_{\mu\nu}(q)\tilde{F}^{a}_{\mu\nu}(-q),
\ee
where we use the notation $\tilde{F}_{\mu\nu}$ for instanton field~(\ref{eq_inst_f}).

One can approximate $\Pi_0(q^2)$ in the whole Euclidean region by~\cite{Agas_Sim_95,Agasian_96,Simonov_93_2}
\be
\Pi_0(q^2) \simeq\frac{11 N_c}{48\pi^2}
\ln \frac{q^2+m^2_*}{\Lambda^2},
\ee
where $m_*$ is connected to $m_0$ through a relation
\be
m^2_*=m^2_0e^{-C}\approx 0,56 m^2_0,~~
C=-\Psi (1) = 0,577 ...
\ee
Inserting in~(\ref{eq_seff_pt_alt_2})
$\tilde{F}^a_{\mu\nu}(q)=-8\pi^2\eta^a_{\mu\nu}\rho^2K_0(\rho q)$,
where $K_0(z)$ is the McDonald function yields
\be
S^{\PP}_{\eff}(\rho)=24\pi^2\int^{\infty}_0
x^3dxK^2_0(x)\frac{1}{g^2(x)}\simeq\frac{8\pi^2}{g^2(1)}
\ee
where we have defined the effective charge
\be
\frac{1}{g^2(x)}= \frac{b}{16\pi^2} \ln
\frac{x^2+m^2_*\rho^2}{\Lambda^2\rho^2}, ~~ b=\frac{11}{3} N_c,
\ee
which is obviously the same result as~(\ref{eq_seff_pt_final}).

\newpage
\section{Bilocal correlator}
\label{appndx_bilocal}

The commonly used parametrization of bilocal correlator~\cite{Dosch_87,DShS,DiG,DiGiacomo_2000}
is different from~(\ref{eq_bilocal}). It has the form
\be
\label{eq_bilocal_2}
\ba
&\langle \tr G_{\mu\nu}(x) \Phi(x,y) G_{\rho\sigma}(y)\Phi(y,x)\rangle \sim
(\delta_{\mu\rho}\delta_{\nu\sigma}-\delta_{\mu\sigma}\delta_{\nu\rho})
\left[\tilde D(z)+\tilde D_1(z)\right]\\
&+(z_{\mu}z_{\rho}\delta_{\nu\sigma}+z_{\nu}z_{\sigma}\delta_{\mu\rho}-
z_{\mu}z_{\sigma}\delta_{\nu\rho}-z_{\nu}z_{\rho}\delta_{\mu\sigma})
\frac{\partial \tilde D_1}{\partial z^2}
\ea
\ee
Clearly, one can establish correspondence between our definition and~(\ref{eq_bilocal_2})
\be
\ba
&\tilde D(z)+\tilde D_1(z)\to D(z)\\
&\frac{1}{2} z^2 \frac{\partial \tilde D_1}{\partial z^2}\to \overline{D}(z)
\ea
\ee
Correlator~(\ref{eq_bilocal_2}) was measured on the lattice~\cite{DiG}. Functions
$\tilde D(z)$ and $\tilde D_1(z)$ were approximated as
\be
\label{eq_d_d1}
\ba
&\tilde D(z)=A_0 \exp\left(-|z|/T_g\right)+\frac{a_0}{z^4}\exp\left(-|z|/\lambda\right)\\
&\tilde D_1(z)=A_1 \exp\left(-|z|/T_g\right)+\frac{a_1}{z^4}\exp\left(-|z|/\lambda\right)
\ea
\ee
Second term in the right hand side of equations~(\ref{eq_d_d1}) corresponds to perturbative
contribution. Numerical values for $T_g$, obtained in~\cite{DiG}, are presented in the Table~\ref{tab_digiacomo}.
Moreover, it was shown in these works that ratio $A_1/A_0$ is a small quantity,
$A_1/A_0 \simeq 0.1$. Turning back to our notation ($D$ and $\overline{D}$), we get for
nonperturbative part
\be
\ba
&D(z)=(A_0+A_1) e^{-z/T_g}\\
&\overline{D}(z) = -\frac{1}{4} \frac{z}{T_g} A_1 e^{-z/T_g}
\ea
\ee
It is clear that in the region $z\sim T_g$, which gives the main contribution to all quantities,
calculated in Section~\ref{sec_IRstab}, $\overline{D}(z)\ll D(z)$ because $A_1 \ll A_0$.

In our calculations we use gaussian parametrization
\be
\ba
&\tilde D(z)=A_0e^{-z^2/T_g^2}\\
&\tilde D_1(z)=A_1e^{-z^2/T_g^2}
\ea
\ee
Thus, we have
\be
\label{eq_D_bar_D}
\ba
&D(z)=(A_0+A_1) e^{-z^2/T_g^2}\\
&\overline{D}(z) = -\frac{1}{2} \frac{z^2}{T_g^2} A_1 e^{-z^2/T_g^2}
\ea
\ee
and still $\overline{D}(z)\ll D(z)$ in the region $z\sim T_g$.

\newpage
\section{Dependence of $S_{\dia}$ on functions $D$ and $\bar D$}
\label{appndx_bar_d}

Let us consider numerical changes in effective action due to taking into
account function $\overline{D}(z)$ in bilocal correlator~(\ref{eq_bilocal}).
As stated above, nonlocal ''diamagnetic'' interaction leads to instanton
stabilization in size. Correspondingly, we will consider changes of
$S_{\dia}$ due to $\overline{D}$ and to taking exponential parametrization
of bilocal correlator. Starting from (\ref{eq_seff_component_2_a})-(\ref{eq_seff_component_2_d})
and taking into account ~(\ref{eq_D_bar_D}) one gets for $S_{\dia}$ (instead of ~(\ref{eq_component_4_a})):

\be
\ba
\label{eq_sdia_alt}
&S_{\dia}=\frac{2\pi^2}{g^2} \frac{\langle
G^2\rangle}{\mu^4}\frac{N_c}{N_c^2-1} \zeta^4
\int\limits_0^{\infty}dx
\frac{x^3}{(x^2+\zeta^2)^2} \varphi(x),\\
&\varphi(x)=\int\limits_0^1 \alpha d \alpha \int\limits_0^1 \beta
d \beta \left[(A_0+A_1)e^{-(\alpha-\beta)^2 x^2} -
  A_1 (\alpha-\beta)^2 x^2 e^{-(\alpha-\beta)^2 x^2}\right],\\
&A_0\simeq 0.9,\quad A_1\simeq 0.1
\ea
\ee

\begin{figure}[!htb]
\begin{picture}(288,188)
\put(0,10){\includegraphics{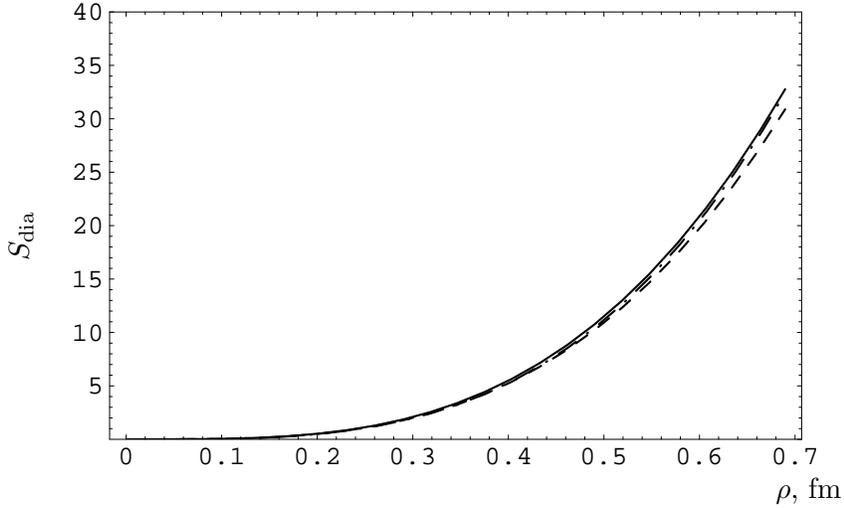}}
\put(270,0){$\rho$, fm}
\put(-20,90){\rotatebox{90}{$S_{\dia}$}}
\end{picture}
\caption{$S_{\dia}$, with  (dashed line) and without (solid line) taking $\overline{D}(z)$ into account.
  Dash-dotted line -- $S_{\dia}$ at $D(z)=\exp\{-|z|/T_g\}$.}
\label{fig_s_dia}
\end{figure}

Fig.~\ref{fig_s_dia} shows $S_{\dia}$ calculated according to
formulas~(\ref{eq_component_4_a})-(\ref{eq_varphi}) and~(\ref{eq_sdia_alt}).
One can see, that the difference is very small, and therefore our approximation,
when we neglect $\overline{D}(z)$ as compared to $D(z)$ is justified.
In the same figure there also is shown graph for $S_{\dia}$, calculated with
$D(z)=e^{-|z|/T_g}$,  $\overline{D}(z)=0$. It is clear that exact form of
function $D(z)$ does not significantly influence the result. The only property
this function should have is to be monotone decreasing function of $z$ with characteristic
correlation length $T_g$.

\vspace{2cm}


\newpage
\begin{thebibliography}{99}

\bibitem{BPST}
A.~M.~Polyakov,
Phys.\ Lett.\ {\bf B59}, 79 (1975);~
%
A.~A.~Belavin, A.~M.~Polyakov, A.~S.~Shvarts and Y.~S.~Tyupkin,
Phys.\ Lett.\ {\bf B59}, 85 (1975).

\bibitem{tHooft_76}
G.~'t~Hooft,
Phys.\ Rev.\ {\bf D14}, 3432 (1976).

\bibitem{CDG_76}
C.~G.~Callan, R.~F.~Dashen and D.~J.~Gross,
Phys.\ Lett.\ {\bf B63}, 334 (1976).

\bibitem{CDG_78}
C.~G.~Callan, R.~F.~Dashen and D.~J.~Gross,
Phys.\ Rev.\ {\bf D17}, 2717 (1978);~
%
Phys.\ Rev.\ {\bf D19}, 1826 (1979).


\bibitem{tHooft_76b}
G.~'t~Hooft,
Phys.\ Rev.\ Lett.\  {\bf 37}, 8 (1976).

\bibitem{Witten_Venez_79}
E.~Witten,
Nucl.\ Phys.\ {\bf B149}, 285 (1979);~
%
G.~Veneziano,
Nucl.\ Phys.\ {\bf B159}, 213 (1979).

\bibitem{Diak_Pet_86}
D.~Diakonov and V.~Y.~Petrov,
Nucl.\ Phys.\ {\bf B272}, 457 (1986).

\bibitem{GI}
B.~V.~Geshkenbein and B.~L.~Ioffe,
Nucl.\ Phys.\ {\bf B166}, 340 (1980).

\bibitem{Scha_Shur_98}
T.~Schafer and E.~V.~Shuryak,
Rev.\ Mod.\ Phys.\  {\bf 70}, 323 (1998) [hep-ph/9610451].

\bibitem{Bali:1999hx}
G.~S.~Bali,
Nucl.\ Phys.\ Proc.\ Suppl.\  {\bf 83}, 422 (2000)
[hep-lat/9908021];~
%
Phys.\ Rept.\  {\bf 343}, 1 (2001)
[hep-ph/0001312].

\bibitem{ShevSim}
Yu.~A.~Simonov,
JETP Lett.\  {\bf 71}, 127 (2000)
[hep-ph/0001244];~
%
V.~I.~Shevchenko and Yu.~A.~Simonov,
Phys.\ Rev.\ Lett.\  {\bf 85}, 1811 (2000)
[hep-ph/0001299].

\bibitem{Fukushima:2000ix}
M.~Fukushima, E.-M.~Ilgenfritz and H.~Toki,
Phys.\ Rev.\ {\bf D64}, 034503 (2001)
[hep-ph/0012358].

\bibitem{Ilgenfritz:1980vj}
E.-M.~Ilgenfritz and M.~Muller-Preussker,
Nucl.\ Phys.\ {\bf B184}, 443 (1981).

\bibitem{Shuryak_81}
E.~V.~Shuryak,
Nucl.\ Phys.\ {\bf B203}, 93 (1982).

\bibitem{Shuryak:nr}
E.~V.~Shuryak,
Phys.\ Rept.\  {\bf 115}, 151 (1984).


\bibitem{Diak_Pet_84}
D.~Diakonov and V.~Y.~Petrov,
Nucl.\ Phys.\ {\bf B245}, 259 (1984).

\bibitem{BY}
I.~I.~Balitsky and A.~V.~Yung,
Phys.\ Lett.\ {\bf B168}, 113 (1986);~
%
A.~V.~Yung,
Nucl.\ Phys.\ {\bf B297}, 47 (1988).

\bibitem{SVZ}
M.~A.~Shifman, A.~I.~Vainshtein and V.~I.~Zakharov,
Nucl.\ Phys.\ {\bf B163}, 46 (1980).

\bibitem{Agas_Sim_95}
N.~O.~Agasian and Yu.~A.~Simonov,
Mod.\ Phys.\ Lett.\ {\bf A10}, 1755 (1995).

\bibitem{Agasian_96}
N.~O.~Agasian,
Phys.\ Atom.\ Nucl.\  {\bf 59}, 297 (1996).

\bibitem{AF_2001}
N.~O.~Agasian and S.~M.~Fedorov,
JHEP {\bf 0112}, 019 (2001)
[hep-ph/0111208];~
N.~O.~Agasian and S.~M.~Fedorov,
hep-ph/0111305.

\bibitem{Abbot_81}L.~F.~Abbot, Nucl.\ Phys. {\bf B184}, 189 (1981).

\bibitem{Polyakov_87}A.~M.~Polyakov, Gauge Fields and Strings. Harwood: Acad.\ Publ. (1987).

\bibitem{Simonov_93_2}
Yu.~A.~Simonov,
Phys.\ Atom.\ Nucl.\  {\bf 58}, 107 (1995)
[hep-ph/9311247];~
Yu.~A.~Simonov and J.~A.~Tjon,
Annals Phys.\  {\bf 300}, 54 (2002)
[hep-ph/0205165].

\bibitem{Dosch_87}
H.~G.~Dosch,
Phys.\ Lett.\ {\bf B190}, 177 (1987);~
%
H.~G.~Dosch and Yu.~A.~Simonov,
Phys.\ Lett.\ {\bf B205}, 339 (1988);~
%
Yu.~A.~Simonov,
Nucl.\ Phys.\ {\bf B307}, 512 (1988).

\bibitem{DShS}
A.~Di~Giacomo, H.~G.~Dosch, V.~I.~Shevchenko and Yu.~A.~Simonov,
hep-ph/0007223.


\bibitem{DiG}
M.~Campostrini, A.~Di~Giacomo and G.~Mussardo,
Z.\ Phys.\ {\bf C25}, 173 (1984);~
%
A.~Di~Giacomo and H.~Panagopoulos,
Phys.\ Lett.\ {\bf B285}, 133 (1992);~
%
A.~Di~Giacomo, E.~Meggiolaro and H.~Panagopoulos,
hep-lat/9603017;~
%
M.~D'Elia, A.~Di~Giacomo and E.~Meggiolaro,
Phys.\ Lett.\ {\bf B408}, 315 (1997) [hep-lat/9705032].

\bibitem{DiGiacomo_2000}
A.~Di~Giacomo,
hep-lat/0012013.


\bibitem{SVZ_79}
M.~A.~Shifman, A.~I.~Vainshtein and V.~I.~Zakharov,
Nucl.\ Phys.\ {\bf B147}, 385,~448 (1979).

\bibitem{BaliBrambillaVairo_98}
G.~S.~Bali, N.~Brambilla and A.~Vairo,
Phys.\ Lett.\ {\bf B421}, 265 (1998)
[hep-lat/9709079].


\bibitem{23}
Yu.~A.~Simonov,
Few Body Syst.\  {\bf 25}, 45 (1998) [hep-ph/9712248];~
%
Yu.~A.~Simonov,
Phys.\ Atom.\ Nucl.\  {\bf 61}, 855 (1998).

\bibitem{NN}
S.~Narison,
Phys.\ Lett.\ {\bf B387}, 162 (1996) [hep-ph/9512348].

\bibitem{IH}
Kerzon Huang, Statistical Mechanics, Jon Wiley and Sons, Inc., New
York--London (1963);~
%
A.~Isihara, Statistical Physics, Academic Press, New York--London
(1971).

\bibitem{MAK}
A.~B.~Migdal, N.~O.~Agasian and S.~B.~Khokhlachev,
JETP Lett.\  {\bf 41}, 497 (1985);~
%
N.~O.~Agasian and S.~B.~Khokhlachev,
Sov.\ J.\ Nucl.\ Phys.\  {\bf 55}, 628,~633 (1992);~
%
N.~O.~Agasian,
hep-ph/9803252;~
%
hep-ph/9904227.

\bibitem{Dor}
A.~E.~Dorokhov, S.~V.~Esaibegian, A.~E.~Maximov and
S.~V.~Mikhailov,
Eur.\ Phys.\ J.\ {\bf C13}, 331 (2000) [hep-ph/9903450].


\bibitem{Hasen_Niet_98}
A.~Hasenfratz and C.~Nieter,
Phys.\ Lett.\ {\bf B439}, 366 (1998) [hep-lat/9806026].

\bibitem{Polikarpov_87}
M.~I.~Polikarpov and A.~I.~Veselov,
Nucl.\ Phys.\ {\bf B297}, 34 (1988).


\bibitem{Latt_conf}
J.~W.~Negele,
Nucl.\ Phys.\ Proc.\ Suppl.\  {\bf 73}, 92 (1999)
[hep-lat/9810053];~
%
M.~Teper,
Nucl.\ Phys.\ Proc.\ Suppl.\  {\bf 83}, 146 (2000)
[hep-lat/9909124];~
%
M.~Garcia Perez,
Nucl.\ Phys.\ Proc.\ Suppl.\  {\bf 94}, 27 (2001)
[hep-lat/0011026].

\bibitem{Ringwald_99}
A.~Ringwald and F.~Schrempp,
Phys.\ Lett.\ {\bf B459}, 249 (1999)
[hep-lat/9903039].


\end{thebibliography}
\end{document}